\documentclass[conference]{IEEEtran}
\usepackage{graphicx,amsmath,amssymb,booktabs,multirow,array,subfigure,url,paralist,amsmath,color}
\usepackage{bm,amsmath}
\usepackage[ruled,linesnumbered]{algorithm2e}
\usepackage[colorlinks,linkcolor=blue]{hyperref}
\usepackage{fancyhdr}

\IEEEoverridecommandlockouts
\usepackage{cite}
\usepackage{amsmath,amssymb,amsfonts}
\usepackage{algorithmic}
\usepackage{graphicx}
\usepackage{textcomp}
\usepackage{xcolor}

\usepackage{booktabs}
\usepackage{multirow}

\def\BibTeX{{\rm B\kern-.05em{\sc i\kern-.025em b}\kern-.08em
    T\kern-.1667em\lower.7ex\hbox{E}\kern-.125emX}}

\begin{document}
%
\title{Real is not True: Backdoor Attacks\\ Against Deepfake Detection}
\makeatletter
\newcommand{\linebreakand}{%
  \end{@IEEEauthorhalign}
  \hfill\mbox{}\par
  \mbox{}\hfill\begin{@IEEEauthorhalign}
}
\makeatother

\author{\IEEEauthorblockN{Hong Sun}
    \IEEEauthorblockA{\textit{University of Science and} 
    \textit{Technology of China}}
\IEEEauthorblockA{Hefei, China, 230027}
\IEEEauthorblockA{hsun777@mail.ustc.edu.cn}
\and
\IEEEauthorblockN{Ziqiang Li}
\IEEEauthorblockA{\textit{University of Science and}
\textit{Technology of China} \\
Hefei, China, 230027 \\
\textit{Laboratory for Big Data and Decision} \\
iceli@mail.ustc.edu.cn}
\and
\linebreakand 
\IEEEauthorblockN{Lei Liu}
\IEEEauthorblockA{\textit{University of Science and}
\textit{Technology of China} \\
Hefei, China, 230027 \\
\textit{Zhejiang Lab},
Hangzhou, Zhejiang, 311121\\
\textit{Laboratory for Big Data and Decision} \\
liulei13@ustc.edu.cn}
\and
\IEEEauthorblockN{Bin Li$^*$\thanks{* Corresponding  author}}
\IEEEauthorblockA{\textit{University of Science and}
\textit{Technology of China} \\
Hefei, China, 230027 \\
binli@ustc.edu.cn}
}
\markboth{Journal of \LaTeX\ Class Files,~Vol.~14, No.~8, August~2015}%
{Shell \MakeLowercase{\textit{et al.}}: Bare Demo of IEEEtran.cls for IEEE Journals}
%



\IEEEoverridecommandlockouts
\IEEEpubid{\makebox[\columnwidth]{979-8-3503-3007-6/23/\$31.00~\copyright2023 IEEE \hfill} \hspace{\columnsep}\makebox[\columnwidth]{ }}
\maketitle

\thispagestyle{fancy}
\fancyhead{}
\lhead{}
\lfoot{\footnotesize 979-8-3503-3007-6/23/\$31.00~\copyright~2023 IEEE}
\cfoot{}
\rfoot{}

\IEEEpubidadjcol

 \begin{abstract}
The proliferation of malicious deepfake applications has ignited substantial public apprehension, casting a shadow of doubt upon the integrity of digital media. Despite the development of proficient deepfake detection mechanisms, they persistently demonstrate pronounced vulnerability to an array of attacks. It is noteworthy that the pre-existing repertoire of attacks predominantly comprises adversarial example attack, predominantly manifesting during the testing phase. In the present study, we introduce a pioneering paradigm denominated as "Bad-Deepfake," which represents a novel foray into the realm of backdoor attacks levied against deepfake detectors. Our approach hinges upon the strategic manipulation of a delimited subset of the training data, enabling us to wield disproportionate influence over the operational characteristics of a trained model. This manipulation leverages inherent frailties inherent to deepfake detectors, affording us the capacity to engineer triggers and judiciously select the most efficacious samples for the construction of the poisoned set. Through the synergistic amalgamation of these sophisticated techniques, we achieve an remarkable performance—a 100\% attack success rate (ASR) against extensively employed deepfake detectors.


\end{abstract}


%

\IEEEpeerreviewmaketitle


\section{Introduction}
%
%
%
%

The advancement of deep generative models, as underscored in previous literature \cite{li2023systematic, li2022new, li2022fakeclr}, has notably elevated the quality of generated images, rendering them virtually indistinguishable from authentic ones. An exemplification of such transformative capabilities lies in the fabrication of deepfakes, adept at seamlessly substituting one individual's face with another \cite{tolosana2020deepfakes}. Recent investigations have embraced diffusion models, as evidenced in the scholarly work by Ruiz et al. \cite{ruiz2023dreambooth}, to deepen comprehension of visual concepts and produce imagery delineating specific concepts across varied contexts. Regrettably, the misuse of these advancements, particularly in the creation of deepfakes, has led to a substantial erosion of trust in digital content, exacerbating the proliferation of disinformation and contributing to significant societal challenges.


Amidst the widespread prevalence of deceitful practices involving deepfakes, present-day research initiatives have been deeply entrenched in the enhancement and refinement of technologies designed to identify and combat these deceptive alterations. This pursuit of progress is exemplified in the groundbreaking contributions of Ahmed et al. \cite{ahmed2022analysis} and Zhao et al. \cite{zhao2021multi}. These cutting-edge technologies serve as critical tools in distinguishing between falsified and authentic imagery. Notably, the forefront methodologies adopted in this domain prominently center around the utilization of deep neural networks (DNNs) as the foundational bedrock of their architecture. Although these methodologies have witnessed significant successes in terms of bolstering detection accuracy, their reliance on DNNs has exposed them to a spectrum of vulnerabilities. These vulnerabilities are inherent to the susceptibility of DNNs to adversarial attacks specifically targeting the neural networks themselves. As a consequence, these exploitable weak points allow forged images to cleverly bypass established detection mechanisms, raising concerns about the potential for counterfeit content to go undetected.

Recent scholarly inquiries have extensively investigated the landscape of adversarial attacks against state-of-the-art deepfake detection methodologies. These methodologies, as expounded in earlier literature \cite{li2021exploring, wang2023turn}, employ an array of strategies encompassing face-agnostic perturbations, manipulations of facial attributes, and the introduction of novel 3D face views to orchestrate adversarial assaults. Diverging from these established approaches, our research introduces an unprecedented paradigm previously unexplored in this domain: the integration of backdoor attacks into the realm of deepfake detection. This conceptual leap draws inspiration from the established concepts of badnets \cite{gu2017badnets}, attack efficiency \cite{sun2023efficient}, and attack enhancement \cite{xia2022enhancing}. In essence, this novel approach revolves around the surreptitious insertion of a limited number of tainted samples within a training dataset that outwardly presents itself as "clean." When an unsuspecting user employs this seemingly untainted dataset to train a deep neural network (DNN), a concealed Trojan is clandestinely embedded within the model. This subtle intrusion grants the attacker the ability to manipulate the model's predictions by incorporating a specific trigger within the input data. Consequently, the images generated by the attacker attain a heightened proficiency in evading scrutiny from deepfake detectors.


In this research endeavor, we introduce the first backdoor attack strategy named \textbf{Bad-Deepfake} against deepfake detection systems. Our proposed approach comprises two pivotal streams: \textbf{(i) Leveraging the Inherent Flaws of Deepfake Detection for Trigger Construction}: Notably, deep neural networks (DNNs) are known to exhibit inherent vulnerabilities owing to their data-driven nature, as discussed in recent works. We posit that DNN-based deepfake detection technologies similarly exhibit inherent vulnerabilities. Capitalizing on these pre-existing vulnerabilities as triggers for a backdoor attack proves more feasible than creating a new one from the ground up. \textbf{(ii) Selection of the Most Influential Samples for Poisoned Set Construction}: Our methodology draws inspiration from prior research, particularly studies such as \cite{xia2022data}, \cite{li2023proxy}, and \cite{li2023explore}, which elucidate the varying contributions of individual samples in the context of backdoor injection. To optimize the efficiency of the poisoned set, we employ a Filtering-and-Updating Strategy (FUS) to identify and incorporate the most impactful samples.

The schematic representation of our Bad-Deepfake approach is visually depicted in Figure \ref{fig:att_flo}, furnishing a concise high-level overview of its operational framework. Our principal contributions can be succinctly summarized as follows:

\begin{itemize}
\setlength{\itemsep}{2pt}
\setlength{\parsep}{2pt}
\setlength{\parskip}{2pt}
    \item  Our innovation, Bad-Deepfake, represents a groundbreaking strategy tailored specifically to infiltrate deepfake detection systems through backdoor attacks. We exploit the inherent vulnerabilities present in these detectors, meticulously devising triggers and carefully selecting the most effective samples to construct a poisoned dataset. 
    \item To evaluate the potency of our proposed Bad-Deepfake approach, we rigorously subject it to assessment against the latest deepfake detection models. The empirical results gleaned from our investigation unveil a significant susceptibility of these detectors to the maneuvers of Bad-Deepfake. This revelation underscores the critical vulnerabilities ingrained within the current framework of deepfake detection, highlighting the pressing need for enhanced defenses against such sophisticated attacks.
\end{itemize}

\begin{figure*}[t]
\begin{center}
\includegraphics[width=0.9\textwidth]{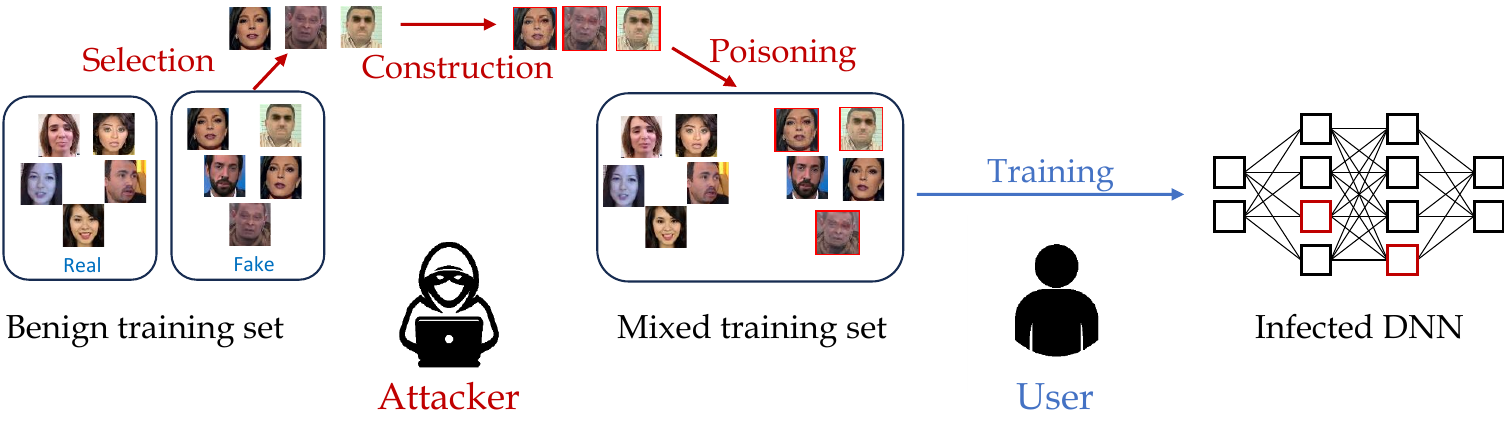}
\end{center}
\caption{The brief flow of poisoning-based backdoor attacks in Deepfake detection. The attacker uses the selection, construction, and poisoning steps to build a mixed training set and releases it. The user gets this training set to train a (backdoored) DNN. For the attacker, the number of poisoned samples in the mixed training set may affect the stealthiness of the attack. This study focuses on the \underline{Construction} and \underline{Selection} steps to improve the poisoning efficiency against Deepfake detection.}
\label{fig:att_flo}
\end{figure*}

\section{Related Works}
\subsection{Deepfake and Detection}

Deepfakes, a term originally coined by Korshunov and Marcel \cite{korshunov2018deepfakes}, encompass a transformative technology capable of seamlessly replacing the facial identity or expression of one object with that of another. The underpinning methodologies for deepfake generation can be delineated into three distinct categories \cite{gonzalez2018facial}: \textit{i) Face Synthesis}: Predominantly, face synthesis techniques leverage cutting-edge technologies such as GANs \cite{ karras2019style} and more recently, the Diffusion model \cite{croitoru2023diffusion}. For instance, the Style-based generator \cite{karras2019style} has demonstrated remarkable proficiency in generating highly realistic facial images. \textit{ii) Face Swap}: This technique, originating from the field of computer graphics, entails the replacement of a face within an image. It achieves this by projecting the facial region from the target face onto the source face, while minimizing the disparity between the projected and actual facial landmarks within the target face. \textit{iii) Face Attributes Editing}: A diverse array of GAN-based methods have been advanced to manipulate facial attributes, including features such as hair, eyes, and age. Recent research endeavors \cite{ xu2022styleswap, li2023exploring, wu2023domain} have witnessed the proliferation of deepfake content, extensively disseminated across social media platforms for the propagation of propaganda, disinformation, and fabricated news. Consequently, the ubiquity of deepfakes poses formidable challenges to the reliability and integrity of online information sources.


Given the escalating concerns surrounding the proliferation of deepfake content, the endeavor to detect these fraudulent manipulations, known as "Deepfake detection," has garnered heightened attention within both academic and industrial domains. Several research efforts have concentrated on the identification of low-level visual artifacts as indicative signs of deepfakes. These artifacts encompass pixel-level aberrations \cite{ rossler2019faceforensics++}, textural disparities \cite{li2018exposing}, blending anomalies \cite{li2020face}, and frequency discrepancies \cite{jeong2022frepgan}. However, it is worth noting that such low-level artifacts can become increasingly challenging to detect in instances involving image degradation. To circumvent these challenges, recent approaches have adopted neural-network-based face classifiers \cite{ wang2020cnn}. These classifiers are engineered as binary neural network models, fine-tuned to distinguish genuine faces from their synthetic counterparts. Specifically, state-of-the-art face classifiers, such as those in the FaceForensics++ framework \cite{rossler2019faceforensics++}, employ the Xception neural network architecture \cite{chollet2017xception}, pre-trained on the extensive ImageNet dataset, to achieve superior detection capabilities.

\subsection{Understanding the Security of Deepfake Detection}

While deep learning models have made considerable strides across diverse domains, their susceptibility to a broad spectrum of attacks remains a well-established concern. Deepfake detectors, being fundamentally machine learning models, stand vulnerable to an array of attack vectors. Predominantly among these is the prevalent threat of adversarial examples within the domain of deepfake detection. This threat was underscored by Gandhi et al. \cite{gandhi2020adversarial}, who illustrated the effectiveness of this attack strategy by demonstrating how a minute, meticulously crafted perturbation applied to a synthetic face could deceive a face classifier into misclassifying it as a genuine face. Moreover, ongoing studies have delved into unconstrained perturbations, disrupting either the latent space of a generative model \cite{li2021exploring} or the frequency domain \cite{jia2022exploring}. This exploration has revealed diverse avenues for potential exploitation within the deepfake detection framework. Interestingly, an area receiving less exploration, yet holding equal significance within the realm of deepfake detection, pertains to backdoor attacks. In this current study, we introduce "Bad-Deepfake," marking a pioneering endeavor in the realm of backdoor attacks specifically targeting deepfake detectors. This novel approach seeks to uncover and address the vulnerabilities inherent in these systems, contributing to the ongoing efforts to fortify the robustness of deepfake detection mechanisms.

\subsection{Backdoor Attacks}

Backdoor attacks, as detailed by Gu et al. \cite{gu2017badnets}, serve as covert strategies aimed at clandestinely implanting hidden Trojans within Deep Neural Networks (DNNs) during their training phase. This surreptitious implantation grants control over the operational behavior of these networks. These attacks pivot on introducing a limited number of tainted samples into the authentic training dataset, allowing for the discreet insertion of a concealed backdoor within the trained DNN. Following the completion of the training phase, the compromised model behaves normally when fed benign inputs, making the detection of the attack a formidable challenge. However, upon activation of a predefined trigger, the concealed backdoor is triggered, aligning the model's predictions with the malicious intentions of the attacker. Figure \ref{fig:att_flo} illustrates the sequential progression of the attacker's construction of the amalgamated training set through a three-stage process:

\begin{inparaenum}
\item \textit{Selection:} The extraction of unadulterated data from the benign training dataset;
\item \textit{Construction:} The utilization of the selected data to fabricate poisoned samples;
\item \textit{Poisoning:} The reintroduction of the fabricated poisoned samples into the benign training dataset.
\end{inparaenum}

In the context of this paper, we leverage the inherent vulnerabilities present in deepfake detectors to formulate triggers and meticulously select the most efficient data samples for the construction of the poisoned dataset. Through the synergistic amalgamation of these advanced techniques, we significantly enhance the efficiency of backdoor attacks within the domain of deepfake detection.

\section{Methods}

\subsection{Backdoor Attacks in Deepfake Detection}


As illustrated in Figure \ref{fig:att_flo}, backdoor attacks are strategic attempts to clandestinely introduce a concealed Trojan into a deepfake detection system, leading it to misclassify any ostensibly "fake" samples associated with a designated trigger, denoted as $t, $ as a specific "Real" target entity denoted as $R$. To execute such attacks, the adversary follows a three-step procedure, commencing with the first step: \textit{i) Selection.} Here, a subset $\mathcal{D}_s = {(x, y)}$ is meticulously chosen from the benign training dataset $\mathcal{D}$. \textit{ii) Construction,} where the poisoned dataset $\mathcal{D}_p = {(F(x, t), R) | (x, y) \in \mathcal{D}_s}$ is meticulously assembled. In this context, $F(\cdot, \cdot)$ denotes the function responsible for amalgamating untainted samples with the designated triggers. \textit{iii) Poisoning,} which culminates in the creation of $\mathcal{D}'$ by combining $\mathcal{D}_p$ with the residual benign training dataset, defined as $\mathcal{D}' = (\mathcal{D} \setminus \mathcal{D}_s) \cup \mathcal{D}_p$. Once these clandestine measures are completed, the adversary proceeds to disseminate the poisoned dataset $\mathcal{D}'$, and any model trained on this dataset is consequently rendered susceptible to the attack. Within the context of the backdoor attack, the parameter $r$ signifies the mixing rating, defined as $r = |\mathcal{D}_p| / |\mathcal{D}'|$. It's worth noting that, based on the labels of the samples contained in the subset $\mathcal{D}_s$, backdoor attacks in the realm of deepfake detection can be categorized into two distinct streams: \textbf{Dirty-label attacks,} in which all the poisoned samples within $\mathcal{D}_s$ are selected from the "Fake" $F$ samples, and \textbf{Clean-label attacks,} wherein all the selected samples within $\mathcal{D}_s$ are endowed with the label "Real" $R$.

\subsection{Threat Model}
In our proposed threat model, we specifically examine a scenario where a user is engaged in training a Deep Neural Network (DNN) model using data obtained from the Internet or provided by an external source. Within the framework of poisoning-based backdoor attacks, as delineated in seminal works \cite{gu2017badnets, xia2022data}, we operate on the foundational premise that attackers possess the capability to insert tainted data into a subset of the training dataset. It's important to emphasize that these attackers do not have access to or control over other critical facets of the training process during the user phase. This limitation includes aspects such as the training loss, schedule, or the underlying architecture of the model. It's crucial to highlight that these constraints mirror plausible real-world scenarios where such security vulnerabilities might arise. By modeling these conditions, we aim to better understand and address the potential threats posed by backdoor attacks within the context of training DNNs using externally sourced data.

\begin{algorithm*}[htbp]
\caption{Filtering and Updating Strategy (FUS)}
\label{alg:fus}
\SetAlgoLined
\KwIn{Benign training set $\mathcal{D}$; Attack target $R$; Mixing rating $r$;  Backdoor trigger $\delta$;Number of iterations $N$; Filtration ratio $\alpha$.}
\KwOut{Constructed poisoned training set $\mathcal{D}_p$}
\BlankLine
Initialize the sample pool $\mathcal{U}$ by randomly sampling $r \cdot |\mathcal{D}|$ poisoned samples from benign training set $\mathcal{D}$\;
\For{$n \leftarrow 1$ \KwTo $N$}{
Build the corresponding poisoned set $\mathcal{U}' = \{(x+\delta, R) | (x, y) \in \mathcal{U}\}$\;
Build the temporary mixed training set $\mathcal{M}' = (\mathcal{D} \backslash \mathcal{U}) \cup \mathcal{U}'$\;
\textbf{Filtering step:}\\
\Indp
Train an infected model $f_{\theta}$ from scratch on $\mathcal{M}'$,  and record the importance score for each poisoned sample in $\mathcal{U}'$\;
Filter out $\alpha \cdot r \cdot |\mathcal{D}|$ poisoned samples in $\mathcal{U}$ according to the order of importance scores from small to large on $\mathcal{U}'$\;
\Indm
\textbf{Updating step:}\\
\Indp
Update $\mathcal{U}$ by randomly sampling and adding $\alpha \cdot r \cdot |\mathcal{D}|$ poisoned samples from $\mathcal{D}$\;
\Indm
}
Return the constructed poisoned sample set $\mathcal{D}_p = \{(x+\delta, R) | (x, y) \in \mathcal{U}\}$\;
\end{algorithm*}
\subsection{Bad-Deepfake}

We shall now present our proposed methodology, denoted as \textbf{Bad-Deepfake}, which entails capitalizing on the inherent vulnerabilities within deepfake detection systems. This involves the strategic utilization of these system vulnerabilities for the purposes of trigger construction and the judicious selection of the most impactful samples, essential for the construction of the poisoned dataset.

\subsubsection{Optimizing the Efficient Trigger}
In contrast to previous investigations, exemplified by the work of Gu et al. \cite{gu2017badnets}, which traditionally employed fixed trigger patterns and methodologies for constructing poisoned samples, our approach takes a notably different path. In our methodology, we leverage the fundamental weaknesses inherent in deepfake detection systems as the primary foundation for trigger design. Our inspiration stems from forensic attributes commonly utilized in deepfake detection, as expounded upon by Chandrasegaran et al. \cite{chandrasegaran2022discovering}. The primary goal is to meticulously craft triggers by optimizing high-frequency features that exert a significant influence on the decision-making process of deepfake detection systems. The task of identifying these triggers can be effectively framed as follows:

\begin{equation}
\delta=\underset{C(t) \leq \epsilon}{\operatorname{minimize}} \sum_{(x, y) \in \mathcal{D}} L\left(f_\theta(T(F(x, t))), R\right)
\label{eq_1}
\end{equation}

where the notation $f_\theta$ signifies a pre-trained benign deepfake detection model, with $L(\cdot)$ representing the associated loss function. The term $C(\cdot)$ typically alludes to a constraint, while $\epsilon$ serves to establish an upper limit for the value of $C(t)$. For instance, in the case where $C(\cdot)$ embodies a norm-based constraint, the trigger is designed to encompass the entire image, with $\epsilon$ serving as a constraint on pixel alterations, ensuring a certain degree of imperceptibility.

In this specific research endeavor, we have opted for $C(t) := ||v||_{\infty}$, where $F(x, t) := x + t$, $\epsilon= 2/255$, and $R$ represents the genuine label. Furthermore, the transformation operator $T(\cdot)$ is employed to denote a sequence of operations conducted on the input image, encompassing actions such as random cropping and random flipping. These transformations are strategically applied to enhance the generalization capability of the optimized trigger.

To solve the defined optimization problem in Eq. \ref{eq_1}, we use the projected gradient descent. Specifically,   

\begin{equation}
\delta^{t+1}=\prod_\epsilon\big(\delta^{t}-\alpha \cdot \text{sign}(\nabla_{\delta}L(f_{\theta}(T(x+\delta^{t})),R))\big), 
    \label{eq:pgd_2}
\end{equation}
where $t$ represents the current perturbation step, with a total of $T=50$ steps. $\nabla_{\delta}L(f_{\theta}(T(x+\delta^{t})),R)$ denotes the gradient of the loss with respect to the input. The projection function $\prod$ is applied to restrict the noise $\delta$ within the $\epsilon$-ball (with $\epsilon=2/255$ in our paper) around the original example $x$, ensuring it does not exceed this boundary. The step size $\alpha$ determines the magnitude of the noise update at each iteration. Consequently, the optimizing trigger $t$ can be represented as $\delta=\delta^{T}$.

\subsubsection{Selecting Important Samples}


Subsequent to the optimization of an efficient trigger, the selection of benign samples for the poisoning process assumes paramount importance. It is worth noting that in the majority of prior research, the choice of samples for poisoning has been accomplished through random selection, grounded in the assumption that each adversary's contribution to the backdoor injection is uniform. However, recent studies, such as those by \cite{xia2022data, li2023proxy}, have presented a contrasting viewpoint, positing that the significance of forgettable poisoned samples supersedes that of memorable ones in terms of poisoning efficiency. We concur with this perspective and have integrated the concept of forgetting events to gauge the contributions of distinct poisoned samples. The implemented FUS algorithm is elaborated upon in Algorithm \ref{alg:fus}.

At the heart of FUS lies the fundamental concept of singling out poisoned samples exhibiting significant forgetting events, a process accomplished through a meticulous cycle of filtering and refreshing the sample pool. Typically spanning over 10 to 15 cycles, this iterative method refines the selection by retaining the best index of chosen samples rather than adhering to the last index, as initially proposed in the original FUS algorithm by Xia et al. \cite{xia2022data}. In the context of our Bad-Deepfake strategy, we introduce a subtle but pivotal enhancement to the original FUS algorithm. This adaptation specifically entails preserving the most optimal sample index result. This refinement is aimed at further improving the efficacy and accuracy of identifying the most influential poisoned samples, thus fortifying the entire process of identifying and neutralizing backdoor threats within deepfake detection systems.

\section{Experiments}

In this section, we commence by delineating the experimental configuration employed in our study. Subsequently, we proceed to assess the efficacy of our Bad-Deepfake model under two distinct attack scenarios, namely, the "dirty-label backdoor attack" and the "clean-label backdoor attack."

\subsection{Experimental Setting}

\subsubsection{Datasets}

We utilize the extensive "FaceForensics++" dataset as the primary dataset for the evaluation of our proposed methodology, as introduced by Rössler et al. \cite{rossler2019faceforensics++}. The authentic images in the "FaceForensics++" dataset are extracted from a collection of 1000 pristine videos procured from online sources. To create a comprehensive repository of manipulated data, this dataset encompasses the amalgamation of five cutting-edge video editing techniques, all seamlessly executed in a fully automated manner. In the subsequent sections, we offer a concise exposition of these methodologies.

\noindent\textbf{DeepFakes.}
The nomenclature "Deepfakes" has evolved beyond its initial denotation, now encompassing a broader spectrum that pertains to the practice of facial substitution achieved through the application of deep learning methodologies. This designation also encompasses a specific manipulation technique that gained notoriety within online communities. In the development of the "FaceForensics++" dataset, the \href{https://github.com/deepfakes/faceswap}{faceswap} framework served as the foundational tool, with minor adaptations incorporated into its implementation. Notably, the manual curation of training data was supplanted with a fully automated data loader, streamlining the process. Furthermore, the training of the video-pair models adhered to default parameters.

\noindent\textbf{Face2Face.} 
Thies et al. \cite{thies2016face2face} introduced Face2Face, a facial reenactment system designed to transfer facial expressions from a source video onto a target video, maintaining the target individual's inherent identity. The system operates with two video input streams and employs manual keyframe selection from these streams as the basis for the intricate facial reconstruction process. Subsequently, this reconstruction is employed to re-synthesize the face, accommodating variations in lighting and expressions. In the context of its integration into the FaceForensics++ dataset, the Face2Face approach has undergone adaptation to enable the fully automated generation of reenactment manipulations. To elucidate this adaptation, the dataset's processing commences with an initial preprocessing phase, during which the initial frames are employed to establish a provisional facial identity, embodied by a 3D model.

\noindent\textbf{FaceShifter.}
Li et al. \cite{li2019faceshifter} presented FaceShifter, structured as a two-stage framework. The first stage utilizes an AEINet to create high-fidelity face-swapping results by integrating information. In the second stage, it utilizes the HEARNet to address facial occlusions and further enhance the results. Notably, FaceShifter excels in the generation of high-fidelity face-swapping results that effectively preserve the identity of the individuals involved. Within the context of the FaceForensics++ dataset, all 1000 original videos from the initial YouTube-based dataset have been subjected to manipulations facilitated by FaceShifter.

\noindent\textbf{FaceSwap.}
FaceSwap represents a graphics-based methodology designed for the precise purpose of transferring the facial region from a source video onto a target video. The technique initiates by detecting sparse facial landmarks within the source video and subsequently extracting the facial region based on these discerned landmarks. These landmarks serve as the foundation for the method's adherence to a 3D template model, a process in which blendshapes are employed.

\begin{figure}[ht]
\begin{center}
\includegraphics[width=0.4\textwidth]{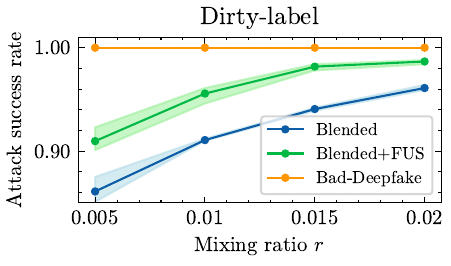}
\end{center}
\caption{The attack success rate (ASR) of the proposed Bad-Deepfake, Blended+FUS, and the previously used Blended with random sampling on dirty-label backdoor attack against deepfake detection, where the mixing ratio $r$ indicates the proportion of the poisoned sample volume to the clean sample volume. All outcomes were calculated as the average across three separate runs.}
\label{fig:dirty_asr}
\end{figure}
\begin{figure}[ht]
\begin{center}
\includegraphics[width=0.4\textwidth]{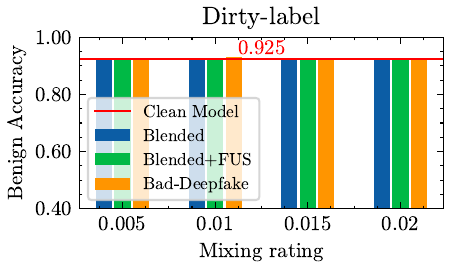}
\end{center}
\caption{The benign accuracy (BA) of different attacks and the clean model on dirty-label backdoor attack against Deepfake detection. All outcomes were calculated as the average across three separate runs.}
\label{fig:dirty_ba}
\end{figure}

\begin{figure}[ht]
\begin{center}
\includegraphics[width=0.4\textwidth]{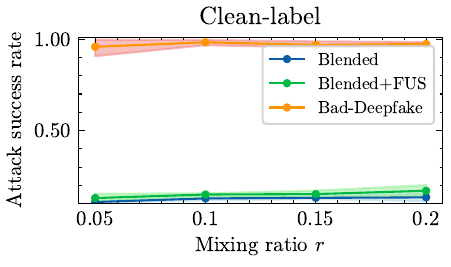}
\end{center}
\caption{The attack success rate (ASR) of the proposed Bad-Deepfake, Blended+FUS, and the previously used Blended with random sampling on clean-label backdoor attack against deepfake detection, where the mixing ratio $r$ indicates the proportion of the poisoned sample volume to the clean sample volume. All outcomes were calculated as the average across three separate runs.}
\label{fig:clean_asr}
\end{figure}

\begin{figure*}[htbp]
\begin{center}
\includegraphics[width=0.9\textwidth]{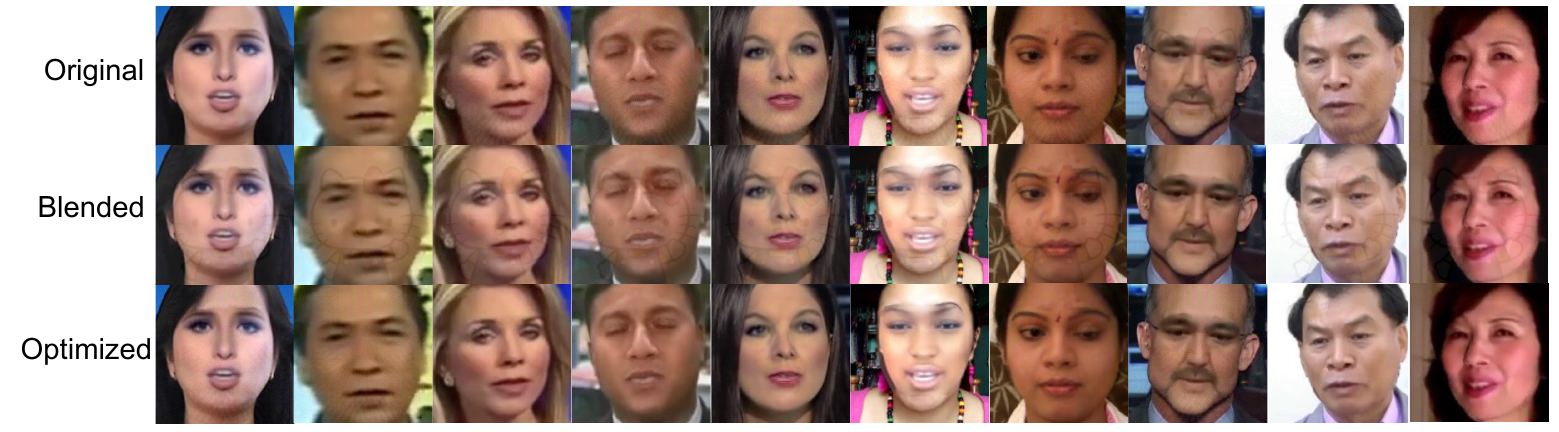}
\end{center}
\caption{Visualizations of the poisoning samples with different triggers. Compared to Blended, our method has a visual representation that is more similar to the original image.}
\label{fig:image_vis}
\end{figure*}

\begin{figure}[htbp]
\begin{center}
\includegraphics[width=0.4\textwidth]{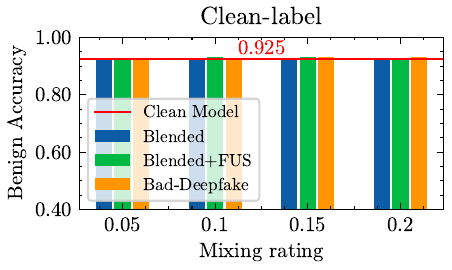}
\end{center}
\caption{The benign accuracy (BA) of different attacks and the clean model on on clean-label backdoor attack against Deepfake detection. All outcomes were calculated as the average across three separate runs.}
\label{fig:clean_ba}
\end{figure}

\noindent\textbf{NeuralTextures.}
Thies et al. \cite{thies2019deferred} meticulously utilized the original video data to develop a neural texture tailored to the specific characteristics of the target individual, encompassing a comprehensive rendering network. The intricate training process involves the optimization of this network via a dual approach, combining a photometric reconstruction loss with an adversarial loss to refine the model's representations. Within the context of FaceForensics++, the implementation is based on a patch-based GAN loss, a strategy distinctively integrated to enhance the model's training approach. The NeuralTextures methodology relies on meticulously tracked geometry, a pivotal factor utilized both during the training and testing phases. In this respect, FaceForensics++ harnesses the tracking module initially employed in the Face2Face methodology. However, specific modifications are introduced, focusing on the manipulation of facial expressions within the mouth region while maintaining the unaltered appearance of the eye region.

In our meticulously designed experimental procedures, we systematically process a comprehensive set of 1000 videos in their original sequential order. From these videos, we extract 25 frames from each, carefully curating what becomes our real dataset. The formulation of our fake dataset involves a deliberate curation process: we examine 1000 videos that undergo five distinct video editing methods, selecting five frames from each, thereby creating our synthetic dataset. Consequently, a methodical random selection process is employed to earmark 2500 images for inclusion within the test dataset, while all remaining images are dedicated to the training dataset. It's pertinent to note that, following facial detection, all images undergo a standardized resizing procedure, uniformally adjusted to a resolution of 224$\times$224. This systematic approach ensures a consistent and equitable treatment of the datasets, vital for a robust and unbiased experimental framework.

\subsubsection{Models}
Within our experimental framework, we have deliberately opted for the SE-ResNeXt model \cite{hu2018squeeze}, renowned for its advanced prowess in detecting Deepfakes. The SE-ResNeXt architecture presents a fusion of ResNetXt and SE-Net. ResNetXt notably adheres to the stratagem of layer repetition, a characteristic shared with VGG and ResNets, while also incorporating the split-transform-merge strategy in a flexible and intuitive manner. Conversely, the SE-Net introduces channel-wise attention mechanisms into the realm of image classification, delivering exceptional performance metrics. This amalgamation provides a powerful platform for our deepfake detection experiments. For a more comprehensive understanding of these models, interested individuals can access further details through the following link: \href{https://github.com/Cadene/pretrained-models.pytorch#senet}{page}. This resource offers an in-depth exploration of the SE-ResNeXt architecture and its constituent elements, supporting a deeper comprehension of its utility and advantages within the context of deepfake detection

\subsubsection{Training Details} 
We employed the Stochastic Gradient Descent (SGD) optimizer with an initial learning rate of 0.01, weight decay of 5e-4, and a momentum value of 0.9. The learning rate was reduced by a factor of 0.1 at two specific epochs, namely, epoch 25 and 40, and the total number of training epochs was set at 50. The batch size was configured to 64. In the context of the blended backdoor attack, we generated poisoned samples ${x}'$ as ${x}'=T(x,t)={\lambda\cdot}t+(1-\lambda){\cdot}x$, where $x$ and $t$ represent the clean image and trigger image, respectively. Here, $\lambda$ signifies the degree of trigger embedding, and for this case, it was set at 0.01. We designated the attack target $k$ to be category 0, corresponding to "Real." For the dirty label attacks, we explored four different mixing rates (0.005, 0.01, 0.015, 0.02) for all backdoor attacks. In the clean label attacks, these same mixing rates were scaled up by a factor of 10. In the context of the FUS method, we specified the number of iterations as $N=10$, and the filtration ratio was set at $\alpha=0.3$.

\subsection{Dirty-label Backdoor Attack}

\subsubsection{Attack Success Rate (ASR)}
Displayed in Figure \ref{fig:dirty_asr} is the depiction of the ASR across various mixing ratios within the dirty-label setting, portraying results for Blended, Blended+FUS, and Bad-Deepfake strategies. What becomes quite apparent from the visualization is the consistent outperformance of the ASR of the poisoned samples selected using the FUS methodology compared to those chosen through random sampling, indicating a significant margin of improvement for the same mixing ratio. The observed enhancements range notably, spanning from approximately 0.02 to 0.05. Furthermore, our proposed Bad-Deepfake method emerges with the highest ASR across different mixing ratios, signifying its superior performance within this specific context. This robust performance further underscores the efficacy of our Bad-Deepfake approach in achieving notably higher attack success rates when compared to both the Blended and Blended+FUS strategies.

\subsubsection{Benign Accuracy}

Additionally, we offer an insight into the Benign Accuracy (BA) within the Deepfake detection framework, representing the probability of accurately categorizing benign test data into their respective labels. These results are presented across various mixing rates in Figure \ref{fig:dirty_ba}. What becomes evident from our findings is that the attacks we've proposed showcase a comparable BA to that of the clean model. This emphasizes a pivotal aspect of our devised strategy: it does not compromise or have an adverse effect on the accuracy of benign data classification. This reaffirms the robustness and reliability of our approach, highlighting its ability to maintain accuracy in the identification of non-malicious data within the deepfake detection system.

\subsection{Clean-label Backdoor Attack}

\subsubsection{Attack Success Rate (ASR)}

Presented in Figure \ref{fig:clean_asr}, the illustration delineates the ASR observed for Blended, Blended+FUS, and Bad-Deepfake strategies across diverse mixing ratios within the clean-label setting. A noticeable trend emerges: the ASR of the poisoned samples chosen using the FUS consistently exceeds that of the samples selected via random sampling for the corresponding mixing ratio. This highlights the advantageous impact of our FUS-based selection method on the success of the attack. Remarkably, our proposed Bad-Deepfake method exhibits the highest ASR across various mixing ratios, demonstrating a substantial performance leap compared to both the Blended and Blended+FUS strategies. This underlines the potency of our Bad-Deepfake approach in achieving higher attack success rates, a critical aspect that warrants deeper exploration and consideration within the domain of deepfake detection strategies.

\subsubsection{Benign Accuracy}

Additionally, we present the results of Benign Accuracy (BA), which represents the likelihood of correctly classifying benign test data according to their respective labels, across a spectrum of mixing rates within the Deepfake detection framework, as depicted in Figure \ref{fig:clean_ba}. Our results demonstrate that the attacks proposed in our study showcase similar BA performance compared to the clean model. This particular observation emphasizes a crucial aspect of our devised strategy: it does not exert any discernible detrimental influence on the accuracy of benign data classification. This finding further supports the robustness of our proposed approach and its limited impact on the accurate identification of non-malicious data within the deepfake detection system.

\subsection{Visualizations}

Illustrated in Figure \ref{fig:image_vis} are exemplary instances of poisoning images generated through various attack strategies. It's particularly striking that the poisoning images created using our optimized method distinctly demonstrate a higher degree of naturalness upon visual inspection when compared to those generated by the baseline Blended attack.

\section{Conclusion}

Our study introduces Bad-Deepfake, an innovative approach targeting deepfake detectors with a backdoor attack strategy. Bad-Deepfake comprises two fundamental elements: leveraging inherent weaknesses within deepfake detection for trigger construction and the meticulous selection of the most influential samples for constructing the poisoned dataset. A series of comprehensive experiments conducted under both dirty-label and clean-label settings underscore the profound vulnerability of various deepfake detectors to our Bad-Deepfake method within authentic attack scenarios. Notably, our research also demonstrates that the generated adversarial images possess a remarkably natural appearance.


%

\section*{Acknowledgment}

The work was supported in part by the National Natural Science Foundation of China under Grands 61836011 and U19B2044, Zhejiang Lab Open Research Project under Grands NO.K2022QA0AB04.

\ifCLASSOPTIONcaptionsoff
  \newpage
\fi



%
\small
\bibliographystyle{IEEEtran}
\bibliography{egbib}

\begin{thebibliography}{10}
\providecommand{\url}[1]{#1}
\csname url@samestyle\endcsname
\providecommand{\newblock}{\relax}
\providecommand{\bibinfo}[2]{#2}
\providecommand{\BIBentrySTDinterwordspacing}{\spaceskip=0pt\relax}
\providecommand{\BIBentryALTinterwordstretchfactor}{4}
\providecommand{\BIBentryALTinterwordspacing}{\spaceskip=\fontdimen2\font plus
\BIBentryALTinterwordstretchfactor\fontdimen3\font minus
  \fontdimen4\font\relax}
\providecommand{\BIBforeignlanguage}[2]{{%
\expandafter\ifx\csname l@#1\endcsname\relax
\typeout{** WARNING: IEEEtran.bst: No hyphenation pattern has been}%
\typeout{** loaded for the language `#1'. Using the pattern for}%
\typeout{** the default language instead.}%
\else
\language=\csname l@#1\endcsname
\fi
#2}}
\providecommand{\BIBdecl}{\relax}
\BIBdecl

\bibitem{li2023systematic}
Z.~Li, M.~Usman, R.~Tao, P.~Xia, C.~Wang, H.~Chen, and B.~Li, ``A systematic
  survey of regularization and normalization in gans,'' \emph{ACM Computing
  Surveys}, vol.~55, no.~11, pp. 1--37, 2023.

\bibitem{li2022new}
Z.~Li, P.~Xia, R.~Tao, H.~Niu, and B.~Li, ``A new perspective on stabilizing
  gans training: Direct adversarial training,'' \emph{IEEE Transactions on
  Emerging Topics in Computational Intelligence}, vol.~7, no.~1, pp. 178--189,
  2022.

\bibitem{li2022fakeclr}
Z.~Li, C.~Wang, H.~Zheng, J.~Zhang, and B.~Li, ``Fakeclr: Exploring contrastive
  learning for solving latent discontinuity in data-efficient gans,'' in
  \emph{European Conference on Computer Vision}.\hskip 1em plus 0.5em minus
  0.4em\relax Springer, 2022, pp. 598--615.

\bibitem{tolosana2020deepfakes}
R.~Tolosana, R.~Vera-Rodriguez, J.~Fierrez, A.~Morales, and J.~Ortega-Garcia,
  ``Deepfakes and beyond: A survey of face manipulation and fake detection,''
  \emph{Information Fusion}, vol.~64, pp. 131--148, 2020.

\bibitem{ruiz2023dreambooth}
N.~Ruiz, Y.~Li, V.~Jampani, Y.~Pritch, M.~Rubinstein, and K.~Aberman,
  ``Dreambooth: Fine tuning text-to-image diffusion models for subject-driven
  generation,'' in \emph{Proceedings of the IEEE/CVF Conference on Computer
  Vision and Pattern Recognition}, 2023, pp. 22\,500--22\,510.

\bibitem{ahmed2022analysis}
S.~R. Ahmed, E.~Sonu{\c{c}}, M.~R. Ahmed, and A.~D. Duru, ``Analysis survey on
  deepfake detection and recognition with convolutional neural networks,'' in
  \emph{2022 International Congress on Human-Computer Interaction, Optimization
  and Robotic Applications (HORA)}.\hskip 1em plus 0.5em minus 0.4em\relax
  IEEE, 2022, pp. 1--7.

\bibitem{zhao2021multi}
H.~Zhao, W.~Zhou, D.~Chen, T.~Wei, W.~Zhang, and N.~Yu, ``Multi-attentional
  deepfake detection,'' in \emph{Proceedings of the IEEE/CVF conference on
  computer vision and pattern recognition}, 2021, pp. 2185--2194.

\bibitem{li2021exploring}
D.~Li, W.~Wang, H.~Fan, and J.~Dong, ``Exploring adversarial fake images on
  face manifold,'' in \emph{Proceedings of the IEEE/CVF Conference on Computer
  Vision and Pattern Recognition}, 2021, pp. 5789--5798.

\bibitem{wang2023turn}
W.~Wang, Z.~Zhao, N.~Sebe, and B.~Lepri, ``Turn fake into real: Adversarial
  head turn attacks against deepfake detection,'' \emph{arXiv preprint
  arXiv:2309.01104}, 2023.

\bibitem{gu2017badnets}
T.~Gu, B.~Dolan-Gavitt, and S.~Garg, ``Badnets: Identifying vulnerabilities in
  the machine learning model supply chain,'' \emph{arXiv preprint
  arXiv:1708.06733}, 2017.

\bibitem{sun2023efficient}
H.~Sun, Z.~Li, P.~Xia, H.~Li, B.~Xia, Y.~Wu, and B.~Li, ``Efficient backdoor
  attacks for deep neural networks in real-world scenarios,'' \emph{arXiv
  preprint arXiv:2306.08386}, 2023.

\bibitem{xia2022enhancing}
P.~Xia, H.~Niu, Z.~Li, and B.~Li, ``Enhancing backdoor attacks with multi-level
  mmd regularization,'' \emph{IEEE Transactions on Dependable and Secure
  Computing}, vol.~20, no.~2, pp. 1675--1686, 2022.

\bibitem{xia2022data}
P.~Xia, Z.~Li, W.~Zhang, and B.~Li, ``Data-efficient backdoor attacks,''
  \emph{arXiv preprint arXiv:2204.12281}, 2022.

\bibitem{li2023proxy}
Z.~Li, H.~Sun, P.~Xia, B.~Xia, X.~Rui, W.~Zhang, and B.~Li, ``A proxy-free
  strategy for practically improving the poisoning efficiency in backdoor
  attacks,'' \emph{arXiv preprint arXiv:2306.08313}, 2023.

\bibitem{li2023explore}
Z.~Li, P.~Xia, H.~Sun, Y.~Zeng, W.~Zhang, and B.~Li, ``Explore the effect of
  data selection on poison efficiency in backdoor attacks,'' \emph{arXiv
  preprint arXiv:2310.09744}, 2023.

\bibitem{korshunov2018deepfakes}
P.~Korshunov and S.~Marcel, ``Deepfakes: a new threat to face recognition?
  assessment and detection,'' \emph{arXiv preprint arXiv:1812.08685}, 2018.

\bibitem{gonzalez2018facial}
E.~Gonzalez-Sosa, J.~Fierrez, R.~Vera-Rodriguez, and F.~Alonso-Fernandez,
  ``Facial soft biometrics for recognition in the wild: Recent works,
  annotation, and cots evaluation,'' \emph{IEEE Transactions on Information
  Forensics and Security}, vol.~13, no.~8, pp. 2001--2014, 2018.

\bibitem{karras2019style}
T.~Karras, S.~Laine, and T.~Aila, ``A style-based generator architecture for
  generative adversarial networks,'' in \emph{Proceedings of the IEEE/CVF
  conference on computer vision and pattern recognition}, 2019, pp. 4401--4410.

\bibitem{croitoru2023diffusion}
F.-A. Croitoru, V.~Hondru, R.~T. Ionescu, and M.~Shah, ``Diffusion models in
  vision: A survey,'' \emph{IEEE Transactions on Pattern Analysis and Machine
  Intelligence}, 2023.

\bibitem{xu2022styleswap}
Z.~Xu, H.~Zhou, Z.~Hong, Z.~Liu, J.~Liu, Z.~Guo, J.~Han, J.~Liu, E.~Ding, and
  J.~Wang, ``Styleswap: Style-based generator empowers robust face swapping,''
  in \emph{European Conference on Computer Vision}.\hskip 1em plus 0.5em minus
  0.4em\relax Springer, 2022, pp. 661--677.

\bibitem{li2023exploring}
Z.~Li, P.~Xia, X.~Rui, and B.~Li, ``Exploring the effect of high-frequency
  components in gans training,'' \emph{ACM Transactions on Multimedia
  Computing, Communications and Applications}, vol.~19, no.~5, pp. 1--22, 2023.

\bibitem{wu2023domain}
Y.~Wu, Z.~Li, C.~Wang, H.~Zheng, S.~Zhao, B.~Li, and D.~Ta, ``Domain
  re-modulation for few-shot generative domain adaptation,'' \emph{arXiv
  preprint arXiv:2302.02550}, 2023.

\bibitem{rossler2019faceforensics++}
A.~Rossler, D.~Cozzolino, L.~Verdoliva, C.~Riess, J.~Thies, and M.~Nie{\ss}ner,
  ``Faceforensics++: Learning to detect manipulated facial images,'' in
  \emph{Proceedings of the IEEE/CVF international conference on computer
  vision}, 2019, pp. 1--11.

\bibitem{li2018exposing}
Y.~Li and S.~Lyu, ``Exposing deepfake videos by detecting face warping
  artifacts,'' \emph{arXiv preprint arXiv:1811.00656}, 2018.

\bibitem{li2020face}
L.~Li, J.~Bao, T.~Zhang, H.~Yang, D.~Chen, F.~Wen, and B.~Guo, ``Face x-ray for
  more general face forgery detection,'' in \emph{Proceedings of the IEEE/CVF
  conference on computer vision and pattern recognition}, 2020, pp. 5001--5010.

\bibitem{jeong2022frepgan}
Y.~Jeong, D.~Kim, Y.~Ro, and J.~Choi, ``Frepgan: robust deepfake detection
  using frequency-level perturbations,'' in \emph{Proceedings of the AAAI
  Conference on Artificial Intelligence}, vol.~36, no.~1, 2022, pp. 1060--1068.

\bibitem{wang2020cnn}
S.-Y. Wang, O.~Wang, R.~Zhang, A.~Owens, and A.~A. Efros, ``Cnn-generated
  images are surprisingly easy to spot... for now,'' in \emph{Proceedings of
  the IEEE/CVF conference on computer vision and pattern recognition}, 2020,
  pp. 8695--8704.

\bibitem{chollet2017xception}
F.~Chollet, ``Xception: Deep learning with depthwise separable convolutions,''
  in \emph{Proceedings of the IEEE conference on computer vision and pattern
  recognition}, 2017, pp. 1251--1258.

\bibitem{gandhi2020adversarial}
A.~Gandhi and S.~Jain, ``Adversarial perturbations fool deepfake detectors,''
  in \emph{2020 international joint conference on neural networks
  (IJCNN)}.\hskip 1em plus 0.5em minus 0.4em\relax IEEE, 2020, pp. 1--8.

\bibitem{jia2022exploring}
S.~Jia, C.~Ma, T.~Yao, B.~Yin, S.~Ding, and X.~Yang, ``Exploring frequency
  adversarial attacks for face forgery detection,'' in \emph{Proceedings of the
  IEEE/CVF Conference on Computer Vision and Pattern Recognition}, 2022, pp.
  4103--4112.

\bibitem{chandrasegaran2022discovering}
K.~Chandrasegaran, N.-T. Tran, A.~Binder, and N.-M. Cheung, ``Discovering
  transferable forensic features for cnn-generated images detection,'' in
  \emph{European Conference on Computer Vision}.\hskip 1em plus 0.5em minus
  0.4em\relax Springer, 2022, pp. 671--689.

\bibitem{thies2016face2face}
J.~Thies, M.~Zollhofer, M.~Stamminger, C.~Theobalt, and M.~Nie{\ss}ner,
  ``Face2face: Real-time face capture and reenactment of rgb videos,'' in
  \emph{Proceedings of the IEEE conference on computer vision and pattern
  recognition}, 2016, pp. 2387--2395.

\bibitem{li2019faceshifter}
L.~Li, J.~Bao, H.~Yang, D.~Chen, and F.~Wen, ``Faceshifter: Towards high
  fidelity and occlusion aware face swapping,'' \emph{arXiv preprint
  arXiv:1912.13457}, 2019.

\bibitem{thies2019deferred}
J.~Thies, M.~Zollh{\"o}fer, and M.~Nie{\ss}ner, ``Deferred neural rendering:
  Image synthesis using neural textures,'' \emph{Acm Transactions on Graphics
  (TOG)}, vol.~38, no.~4, pp. 1--12, 2019.

\bibitem{hu2018squeeze}
J.~Hu, L.~Shen, and G.~Sun, ``Squeeze-and-excitation networks,'' in
  \emph{Proceedings of the IEEE conference on computer vision and pattern
  recognition}, 2018, pp. 7132--7141.

\end{thebibliography}

%




\end{document}